\documentclass[  superscriptaddress, amsmath,amssymb, aps,twocolumn,longbibliography]{revtex4-1} 
\usepackage{graphicx}
\usepackage{dcolumn}
\usepackage{amsthm}
\usepackage{bm}

\pdfoutput=1
\usepackage{graphicx,graphics,epsfig,subfigure,times,bm,bbm,amssymb,amsmath,amsthm,mathrsfs,MnSymbol}
\usepackage{gensymb}
\usepackage{amsfonts}
\usepackage[matrix,frame,arrow]{xypic}
\usepackage[pdfstartview=FitH]{hyperref}
\hypersetup{
    colorlinks=true,       
    linkcolor=red,          
    citecolor=magenta,        
    filecolor=magenta,      
    urlcolor=cyan,           
    runcolor=cyan}
\usepackage{epstopdf}
\usepackage[pdftex]{color}
\usepackage{braket}
\usepackage{enumerate}
\usepackage[normalem]{ulem}
\usepackage[usenames,dvipsnames]{xcolor}
\usepackage{multirow}
\usepackage{mathtools}

\definecolor{orange}{rgb}{1,0.5,0}

\begin{document}

\title{Probing prethermal nonergodicity through measurement outcomes of monitored quantum dynamics} 

\author{Zheng-Hang Sun}
\affiliation{Theoretical Physics \uppercase\expandafter{\romannumeral3}, Center for Electronic Correlations and Magnetism, Institute of Physics, University of Augsburg, D-86135 Augsburg, Germany}

\author{Fabian Ballar Trigueros}
\affiliation{Theoretical Physics \uppercase\expandafter{\romannumeral3}, Center for Electronic Correlations and Magnetism, Institute of Physics, University of Augsburg, D-86135 Augsburg, Germany}

\author{Qicheng Tang}
\affiliation{School of Physics, Georgia Institute of Technology, Atlanta, GA 30332, USA}

\author{Markus Heyl}
\affiliation{Theoretical Physics \uppercase\expandafter{\romannumeral3}, Center for Electronic Correlations and Magnetism, Institute of Physics, University of Augsburg, D-86135 Augsburg, Germany}
\affiliation{Centre for Advanced Analytics and Predictive Sciences (CAAPS), University of Augsburg, Universitätsstr. 12a, 86159 Augsburg, Germany}

\begin{abstract}
\noindent Projective measurements are a key element in quantum physics and enable rich phenomena in monitored quantum dynamics. Here, we show that the measurement outcomes, recorded during monitored dynamics, can provide crucial information about the properties of the monitored dynamical system itself. We demonstrate this for a Floquet model of many-body localization, where we find that the prethermal many-body localized regime becomes unstable against rare measurements, yielding an unusual enhancement of quantum entanglement. Through an unsupervised learning and mutual information analysis on the classical dataset of measurement outcomes, we find that the information loss in the system, reflected by the increased entanglement, is compensated by an emergent structure in this classical dataset. Our findings highlight the crucial role of measurements and corresponding classical outcomes in capturing prethermal nonergodicity, offering a promising perspective for applications to other monitored quantum dynamics. 
\end{abstract}
\pacs{Valid PACS appear here}
\maketitle

Acting in unitary quantum evolution with projective measurements can fundamentally alter the quantum entanglement properties of a system~\cite{PhysRevB.98.205136,PhysRevX.9.031009,PhysRevX.10.041020,PhysRevB.100.134306,PhysRevResearch.2.013022,PhysRevB.101.060301,PhysRevB.101.104301,PhysRevB.101.104302,PhysRevB.102.014315,PhysRevLett.125.070606,PhysRevLett.125.210602,PhysRevB.106.144313,PhysRevLett.128.010604,PhysRevLett.128.130605,PhysRevLett.130.220404,PhysRevLett.131.020401,2023arXiv230202934M,PhysRevB.109.L020304,Noel:2022wr,Koh:2023tb,Hoke:2023tr,2025arXiv250113005H,Paviglianiti2024enhanced,PRXQuantum.5.030329,PhysRevX.13.041046,PhysRevB.105.094303,PhysRevLett.131.060403}, and the classical measurement outcomes can potentially offer insights into its non-equilibrium dynamics~\cite{PhysRevLett.129.200602,PRXQuantum.5.020304,PhysRevX.14.041012,PhysRevB.109.094209,PhysRevB.108.054307}. A prominent example is the measurement-induced phase transition (MIPT) in systems governed by scrambling unitary dynamics, where a disentangled phase characterized by an area-law entanglement entropy (EE) emerges at high measurement rates~\cite{PhysRevB.98.205136,PhysRevX.9.031009,PhysRevX.10.041020,PhysRevB.100.134306,PhysRevResearch.2.013022,PhysRevB.101.060301,PhysRevB.101.104301,PhysRevB.101.104302,PhysRevB.102.014315,PhysRevLett.125.070606,PhysRevLett.125.210602,PhysRevB.106.144313,PhysRevLett.128.010604,PhysRevLett.128.130605,PhysRevLett.130.220404,PhysRevLett.131.020401,2023arXiv230202934M,PhysRevB.109.L020304,Noel:2022wr,Koh:2023tb,Hoke:2023tr,2025arXiv250113005H}. The MIPT can also manifest as learnability transitions, wherein sufficiently frequent measurements can leak out information about the initial state of the system, enabling its retrieval from the classical measurement record~\cite{PRXQuantum.5.020304,PhysRevX.14.041012,PhysRevB.109.094209,PhysRevB.108.054307}. Beyond conventional MIPTs, replacing scrambling unitary dynamics with strictly non-scrambling controlled-Z (CZ) gates introduces a new type of MIPT, where measurements can play an active role in the EE~\cite{PhysRevX.11.011030,PhysRevB.108.094104}. Moreover, the study of MIPT with deeply many-body localized (MBL) unitary evolution~\cite{PhysRevResearch.2.043072}, as an extremely slow scrambling dynamics that occurs in interacting quantum systems with sufficiently strong disorder~\cite{Nandkishore:2015vg,Altman:2018vb,Abanin:2019va,Sierant_2025}, suggests that the nature of MIPTs is dependent on the ergodicity (or nonergodicity) of the system. Generally, the changes of quantum entanglement induced by measurements are intrinsically linked to a transformation in the information content of the underlying quantum states. This naturally raises two key questions: Where does this information flow to or from, and how is the corresponding change connected to the ergodic (or non-ergodic) properties of the system? 

In this work, we study the information content contained in the measurement outcomes recorded during monitored quantum dynamics and show how this information can be utilized to characterize the dynamical properties of the underlying quantum system. For concreteness, we consider the monitored quantum dynamics of a Floquet model of MBL with a broad intermediate region of prethermal MBL at finite system sizes interspersed with rare measurements in the $Z$-basis. We find that this prethermal MBL phase becomes unstable in the presence of the rare measurements yielding an unusual enhancement of quantum entanglement and therefore information loss. We demonstrate that these changes in the entanglement properties transform into information content contained in the dataset of the classical measurement outcomes recorded during the dynamics. In particular, we find by means of a principal component analysis (PCA) that the classical dataset of measurement outcomes becomes more structured (i.e., having a lower effective dimension)~\cite{Pearson:1901vg,MEHTA20191}. We further show that this is accompanied by strong temporal correlations between the measurement outcomes with infinite correlation length, as revealed by a mutual information analysis. Our findings suggest that changes in entanglement properties during monitored dynamics can be compensated by information content contained in the dataset of classical measurement outcomes. 

\begin{figure}[]
	\centering
	\includegraphics[width=1\linewidth]{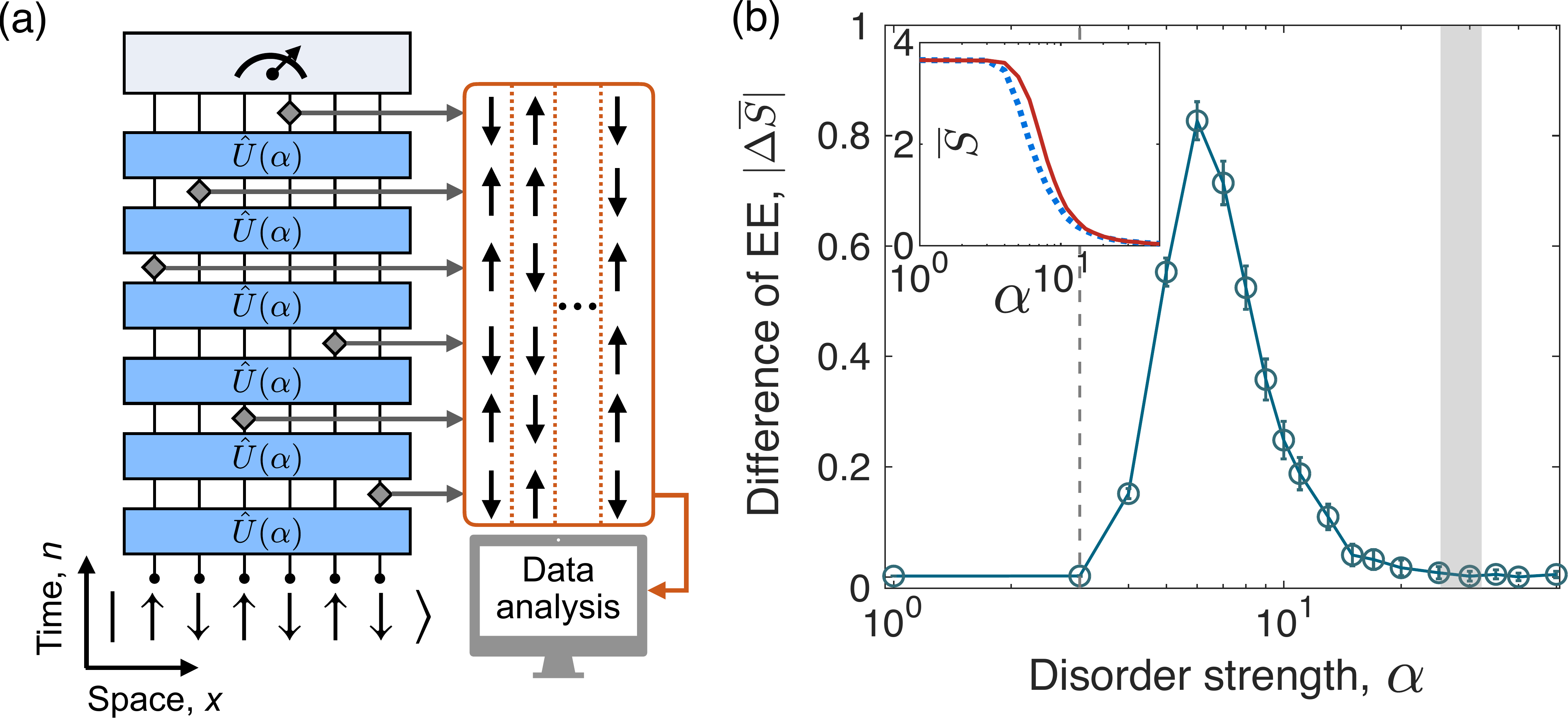}\\
	\caption{(a) Schematic illustration of our monitored quantum circuit, whose unitary dynamics is governed by a Floquet model of MBL with gates $\hat{U}(\alpha)$ interspersed with stroboscopic projective measurements in the \textit{Z} direction. We record the outcome of measurements during the dynamics of the quantum circuit in the form of a dataset, where each column represents the measurement outcomes of a single realization for the monitored quantum circuit. This dataset is then analyzed by means of principal component analysis (PCA) and mutual information. (b) For the system size $L=12$, the difference of the long-time entanglement entropy (EE) between the purely unitary evolution and the quantum circuit interspersed with rare measurements (at probability $p=10^{-4}$), i.e., $|\Delta \overline{S}| = |\overline{S}(p=10^{-4}) - \overline{S}(p=0)|$ as a function of disorder strength $\alpha$. Here, two vertical dashed lines highlight an estimation of the prethermal MBL regime with the lower boundary $\alpha \simeq 3$ and the upper boundary $\alpha \in [25,32]$ based on Ref.~\cite{PhysRevB.105.174205}. The inset shows the long-time EE for the unitary dynamics (dotted line) and with rare measurements (solid line) as a function of disorder strength $\alpha$. }\label{fig1}
\end{figure}

\emph{Quantum circuits.}--In the following we consider a one-dimensional system of $L$ qubits with open boundary conditions. The monitored quantum circuit consists of a unitary evolution described by a Floquet model of many-body localization with unitaries $\hat{U}(\alpha)$ where $\alpha$ denotes the disorder strength, and projective measurements in the \textit{Z} basis through projectors $P_{\pm} = (1\pm Z)/2$ whose outcomes are given by the Born rule [see Fig.~\ref{fig1}(a)]. We will consider these measurements to only occur rarely at a measurement rate $p\simeq 10^{-4}$.

Specifically, the unitary part can be expressed as $\hat{U}(\alpha)=\hat{U}_{u} [\otimes_{i=1}^{L} \hat{d}_{i}]$, which is comprised of single-qubit gates $\hat{d}_{i}$ created by sampling a 2D random matrix from the circular unitary ensemble and subsequently diagonalizing it, and two-qubit gates 
\begin{eqnarray}
\hat{U}_{u} = \otimes_{j=1}^{L-1} \exp[(\frac{i}{\alpha}) M_{k_{j},k_{j}+1}],
\label{ee_linear}
\end{eqnarray}
with $\alpha$ denoting the disorder strength, $M_{k_{j},k_{j}+1}$ sampled from Gaussian unitary ensemble, and $k_{j}\in S_{L-1}$ being a random permutation~\cite{PhysRevB.105.174205}. Without loss of generality, the initial state $|\psi_{0}\rangle=|\uparrow\downarrow\dots\rangle$ for our dynamics is chosen as the N\'{e}el state. The level spacing ratio, as a conventional measure for nonergodicity and therefore MBL, exhibits a crossover point at $\alpha^{(r)}\simeq 6$~\cite{PhysRevB.105.174205}. However, recent studies of the prethermal MBL regime now indicate that the (deep) MBL regime may only occur for much larger disorder strength $\alpha \gtrsim 25$~\cite{10.21468/SciPostPhys.12.6.201,PhysRevB.105.174205,PhysRevLett.131.106301}.

\emph{Dataset of classical measurement outcomes.}--During the monitored quantum dynamics illustrated in Fig.~\ref{fig1}(a), we collect the individual measurement outcomes (either $\uparrow$ or $\downarrow$) in chronological order yielding for each realization of the dynamics a vector \(\textbf{m} = (m_{n_{1}}, \ldots, m_{n_{M}}) \in \{\downarrow, \uparrow\}^M\). Notice, that this construction is unique in our case, because we work in the limit of a very low measurement rate \(p=10^{-4}\) so that practically excludes two measurements at the same time step. The system size is fixed at \(L=12\), and with a final time of \(n_{f} = 5 \times 10^{4}\), the total number of measurements is \(M = 60\).

For the statistical analysis of the monitored quantum circuit in Fig.~\ref{fig1}(a), we generate numerous realizations  $N_{r}$ of the quantum circuit with different random matrices $\hat{d}_{i}$ and $M_{k_{j},k_{j}+1}$, as well as the positions and times of the measurements. We then get one measurement record for each sample $\textbf{m}$. The full dataset of recorded measurements can be represented by the data matrix (see also Fig.~\ref{fig1}(a) for the schematic of dataset)

\begin{equation}
\textbf{M} = 
\begin{bmatrix}
\textbf{m}^{(1)} \\
\textbf{m}^{(2)} \\
\vdots \\
\textbf{m}^{(N_{r})}
\end{bmatrix}
=
\begin{bmatrix}
\downarrow & \uparrow & \cdots & \downarrow \\
\uparrow & \downarrow & \cdots & \uparrow \\
\vdots & \vdots & \ddots & \vdots \\
\downarrow & \uparrow & \cdots & \downarrow
\end{bmatrix}_{N_{r} \times M},
\label{eq_dataset}
\end{equation}
where $\textbf{m}^{(k)}$ denotes the measurement record of $k-$th realization of the quantum circuit. We note that in the further analysis of the dataset, the elements $\uparrow$ ($\downarrow$) in Eq.~(\ref{eq_dataset}) can also be replaced by $1$ ($0$).

\emph{Dynamics of entanglement entropy.}--Before performing the analysis of the classical dataset, let us first discuss our findings for the dynamical properties of the monitored quantum circuit. Specifically, we consider the time evolution of the half-chain  entanglement entropy (EE) defined by $S = -\text{Tr}(\rho_{A}\log\rho_{A})$. Here, the reduced density matrix of the subsystem $A$, encompassing the qubits $i\in\{1,2,...,L/2\}$, is denoted as $\rho_{A} = \text{Tr}_{\overline{A}} |\psi (n) \rangle\langle \psi(n) |$, which can be obtained by tracing out the complement $\overline{A}$, with $|\psi (n) \rangle = [U(\alpha)]^{n}|\psi_{0}\rangle$ being the state at the time step $n$. 

The main results for the long-time dynamics of the EE are summarized in Fig.~\ref{fig1}(b), where we plot the difference of the EE at long times between the purely unitary case and the monitored case included rare measurements. Numerically, we adopt a rate of measurements $p=10^{-4}$, and calculate the dynamics of EE up to time $n\sim 10^{4}$. To minimize temporal fluctuations we perform a slight time average $\overline{S} = [\int_{n_{i}}^{n_{f}} \text{d}n S(n) ]/ (n_{f} - n_{i})$ with the time interval $n\in[n_{i},n_{f}] = [3\times 10^{4}, 5\times 10^{4}]$ for different disorder strengths $\alpha$ and system size $L=12$. We denote the such time-averaged EE with rare measurements and unitary case as $\overline{S}(p=10^{-4})$ and $\overline{S}(p=0)$, respectively. The difference of EE defined as $|\Delta \overline{S}| = |\overline{S}(p=10^{-4}) - \overline{S}(p=0)|$ is shown in Fig.~\ref{fig1}(b), indicating that for both the well-thermalizing phase with $\alpha \lesssim 3$ and deep MBL regime with $\alpha \gtrsim 25$, the rare projective measurements with a small rate $p=10^{-4}$ do not substantially change the EE. Notice that we don't extrapolate the data to the asymptotic infinite-time limit in line with an actual potential experiment where also finite times would only be accessible.

Surprisingly, the case of intermediate values of $\alpha$, which coincides compellingly well with the recently proposed prethermal MBL regime~\cite{PhysRevB.105.174205,PhysRevLett.131.106301}, behaves completely differently. We reveal that the measurements, although vanishingly rare, can significantly increase the EE  [see also the inset of Fig.~\ref{fig1}(b) for a direct comparison with $\overline{S}(p=10^{-4})$ and $\overline{S}(p=0)$]. This indicates an instability of the prethermal MBL regime with two non-commuting limits of $n\rightarrow\infty$ and $p\rightarrow 0$.  

\begin{figure}[]
	\centering
	\includegraphics[width=1\linewidth]{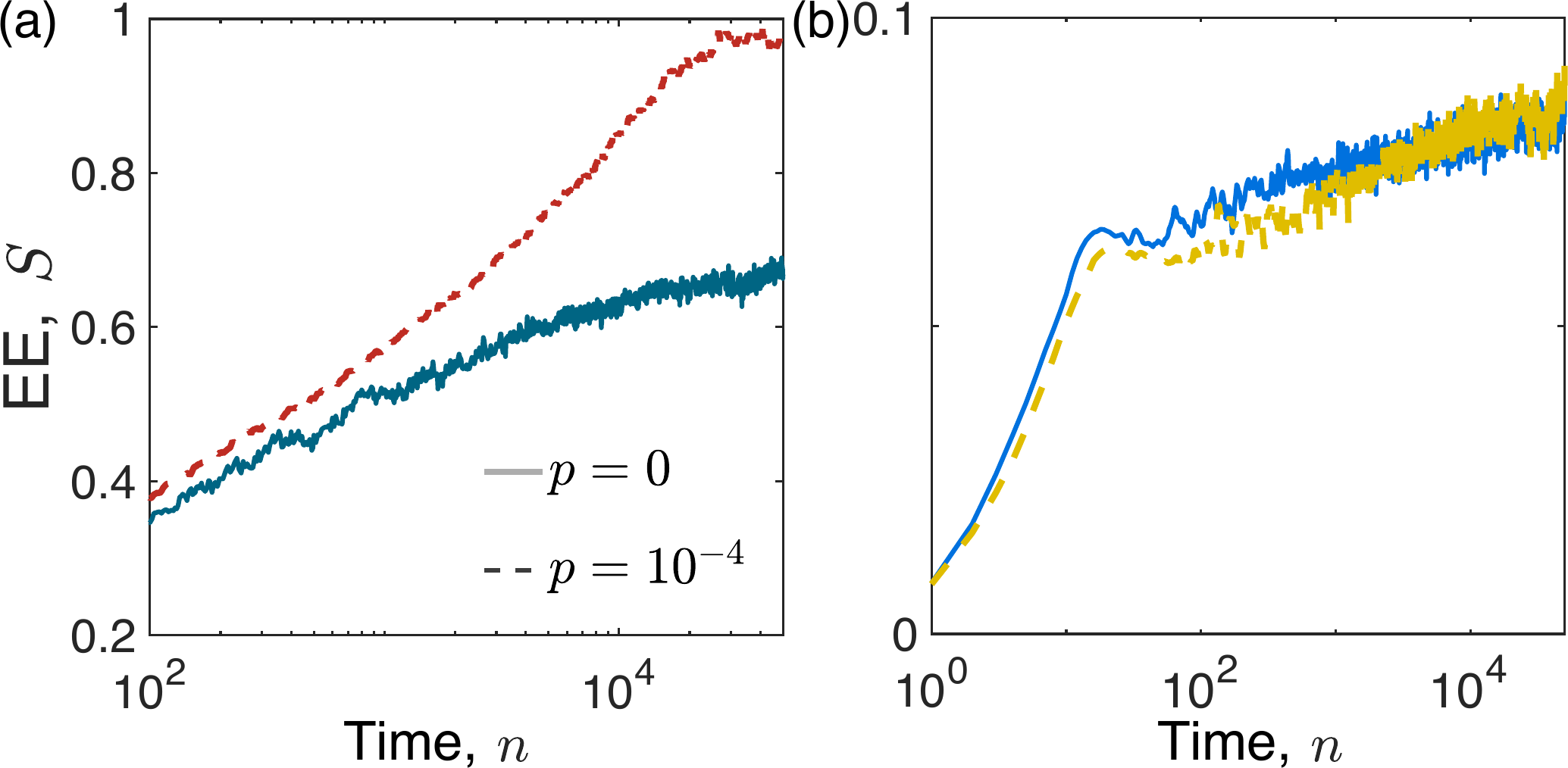}\\
	\caption{Comparison of the entanglement entropy (EE) dynamics for the quantum circuit in Fig.~\ref{fig1}(a) with system size $L=12$ comparing purely unitary evolution (with measurement rate $p=0$) and rare measurements ($p=10^{-4}$) up to a final stroboscopic time $n=5\times 10^{4}$ for disorder strength $\alpha=10$ (a) and $\alpha=30$ (b). The number of realizations for the monitored quantum circuit is larger than $10^{3}$. }\label{fig2}
\end{figure}

We now focus on the dynamics of the EE in detail. As shown in Fig.~\ref{fig2}(a), at an intermediate value of disorder strength $\alpha = 10$, the dynamics of EE is sensitive to rare measurements, which accelerate its growth and lead to a larger long-time value as compared to the unitary case. In contrast, for a stronger disorder $\alpha =30$, Fig.~\ref{fig2}(b) shows that the effect of rare measurements is less pronounced, and in both the unitary ($p=0$) and monitored ($p=10^{-4}$) cases, EE exhibits a slow logarithmic growth, which is similar to the dynamical behaviors of EE in Hamiltonian models of MBL~\cite{PhysRevLett.110.260601,PhysRevLett.109.017202,PhysRevB.93.060201,Singh_2016}. 

Based on the results of EE, one can see that even with a relatively small system size $L=12$, the boundary of the deep MBL regime is at much larger $\alpha$ than that estimated by the level spacing ratio $\alpha^{(r)}\simeq 6$. In the Supplementary Materials, we present the data for larger system sizes, showing similar behaviors, and discuss the finite-size and finite-time effects. 

Several side remarks are in order. First,  if we adopt the forced measurement, so that the outcomes of the measurements are imposed to be $|\downarrow\rangle$ ($|\uparrow\rangle$) for even (odd) sites~\cite{PRXQuantum.2.010352}, the enhancement of EE by rare measurements in the prethermal MBL regime is absent (see numerical results in the Supplementary Materials). This indicates that the Born rule plays a key role in the observed instability, which cannot be attributed solely to the breakdown of time periodicity. Second, the enhancement of EE due to measurements restricted to the $Z$-basis is essentially different from the measurement-only dynamics~\cite{PhysRevX.11.011030}, where the measurements in $X$-basis are necessary to play an active role in the EE. Third, in the Supplementary Materials, we also consider the monitored dynamics of disordered Heisenberg chains, as a standard model for studying many-body localization, with rare measurements. We show that 
deep MBL regime occurs with $W\gtrsim 15$, significantly exceeding the conventional transition point at $W\simeq 3$, and with $W\lesssim15$, the instability induced by rare measurements with enhanced EE is also observed, revealing a similar broad prethermal MBL regime in the Hamiltonian model.

\begin{figure}[]
	\centering
	\includegraphics[width=1\linewidth]{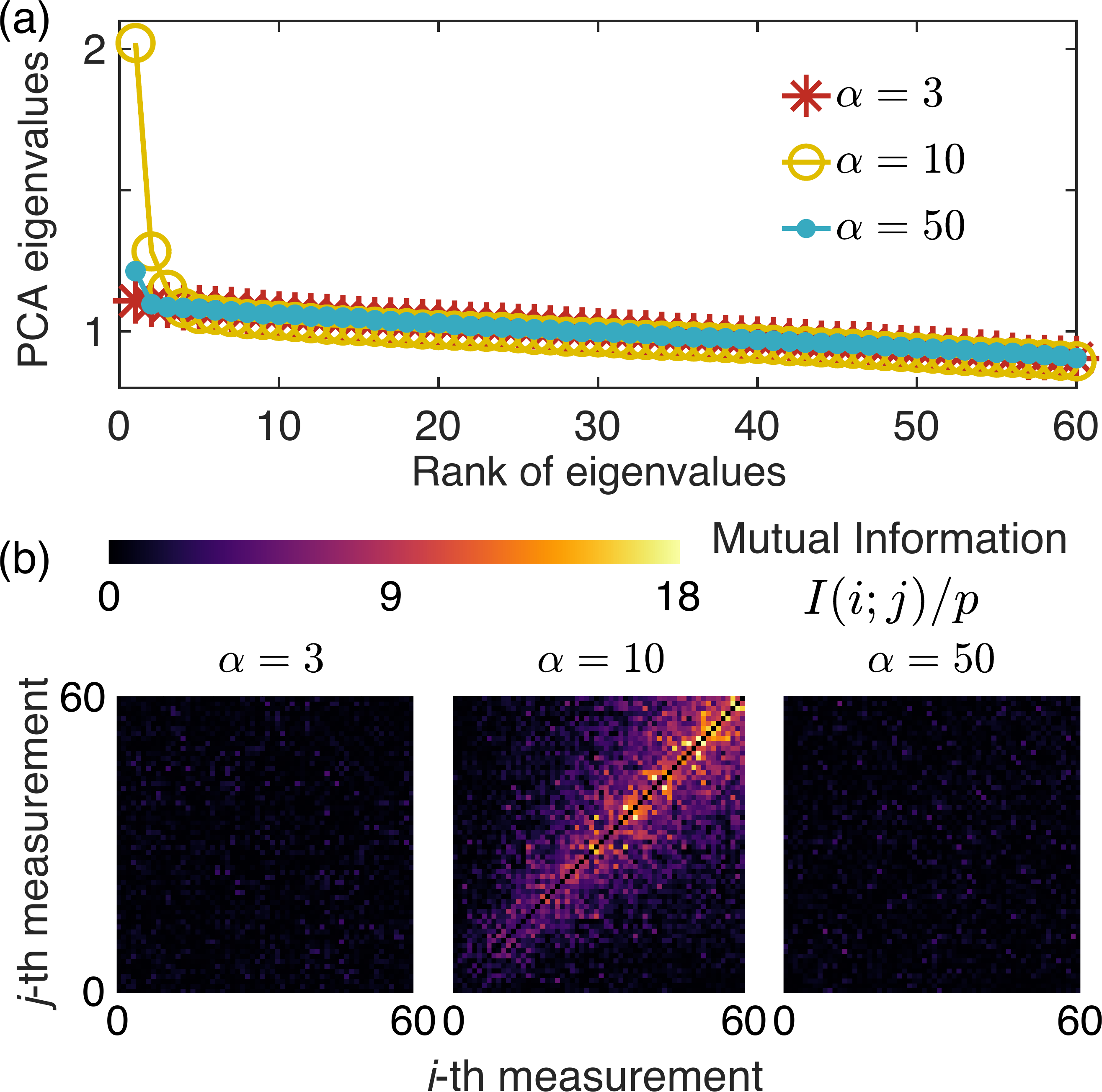}\\
	\caption{(a) The PCA eigenvalues for the measurement dataset extracted from the dynamics of the quantum circuit with system size $L=12$ and three typical values of disorder strength $\alpha=3$, $10$, and $50$. (b) The values of mutual information between the $i-$th and $j-$th measurement outcome $I(i;j)$ for $\alpha=3$, $10$, and $50$. Here, we always consider rare measurements with a rate $p=10^{-4}$.}\label{fig3}
\end{figure}

\emph{Analysis of the classical dataset.} -- The enhancement of EE in the prethermal MBL regime suggests an information loss in the non-equilibrium states due to measurements. It is natural to predict that the information can leak into the measurement record. To demonstrate this viewpoint, in the following, we quantify the information content of the classical measurement outcomes. 

We start by analyzing the dataset (\ref{eq_dataset}) through the PCA, which has been widely employed to detect equilibrium phase transitions~\cite{PhysRevB.94.195105,PhysRevE.96.022140,PhysRevB.96.144432} and MIPTs~\cite{PhysRevB.106.144313,2024arXiv240512863M}. With the $M\times N_{r}$ design matrix $\textbf{M}$ defined by Eq.~(\ref{eq_dataset}), the $M$-dimensional symmetric covariance matrix is 
\begin{eqnarray}
\Sigma (\textbf{M}) = \frac{1}{N_{r}-1} \textbf{M} \textbf{M}^{T}. 
\label{eq_pca}
\end{eqnarray}
By diagonalizing the matrix $\Sigma (\textbf{M})$, we can obtain the eigenvalues of $\Sigma (\textbf{M})$ as the PCA eigenvalues. Here, we consider the number of realizations $N_{r}=3\times 10^{4}$, 
which is sufficient to obtain robust results (see Supplementary Materials for more details). 

We plot the PCA eigenvalues for the dataset for three typical values of disorder strength $\alpha=3$ (thermal regime), $10$ (prethermal MBL regime), and $50$ (deep MBL regime) in Fig.~\ref{fig3}(a) with descending order. Notably, for the prethermal MBL case, the first PCA eigenvalue is significantly larger than others. This clearly shows that there is one dominant principle component, and the dataset has a clear structure. However, for both the thermal and deep MBL regime, the dataset are less structural. 

To characterize this structure of the dataset, we consider the mutual information between the $i$-th and $j$-th measurements at different times, which can quantify the correlations between the measurement outcomes. It is defined as $I(i;j) = D_{\text{KL}}(p_{ij} || p_{i}p_{j})$ with $p_{i (j)}$ being the marginal distribution of the measurement outcomes at time $i$ ($j$), $p_{ij}$ being the joint probability distribution, and $D_{\text{KL}}$ referring to the Kullback-Leibler divergence. A detailed discussion of its computation is provided in the End Matter. 
In Fig.~\ref{fig3}(b), we present the heat maps of the values $I(i;j)/p$ for three typical values of $\alpha$. In the prethermal MBL regime ($\alpha=10$), we observe high values of the mutual information, which suggests strong correlations in the measurement outcomes at nearby times, which are absent in the thermal ($\alpha=3$) and deep MBL regime ($\alpha=50$). 

In the thermal regime, the lack of correlations can be explained by the fact that the information from earlier measurements experiences fast scrambling. Thus, for the rare-measurement limit, the previous measurement does not influence the subsequent one. In the deep MBL regime, the localized nature prevents the information of measurements spreading, leaving the effect of measurements isolated. The strong correlations of subsequent measurement outcomes with $\alpha=10$ reveals a more essential role of rare measurements in the prethermal MBL regime.

\begin{figure}[]
	\centering
	\includegraphics[width=1\linewidth]{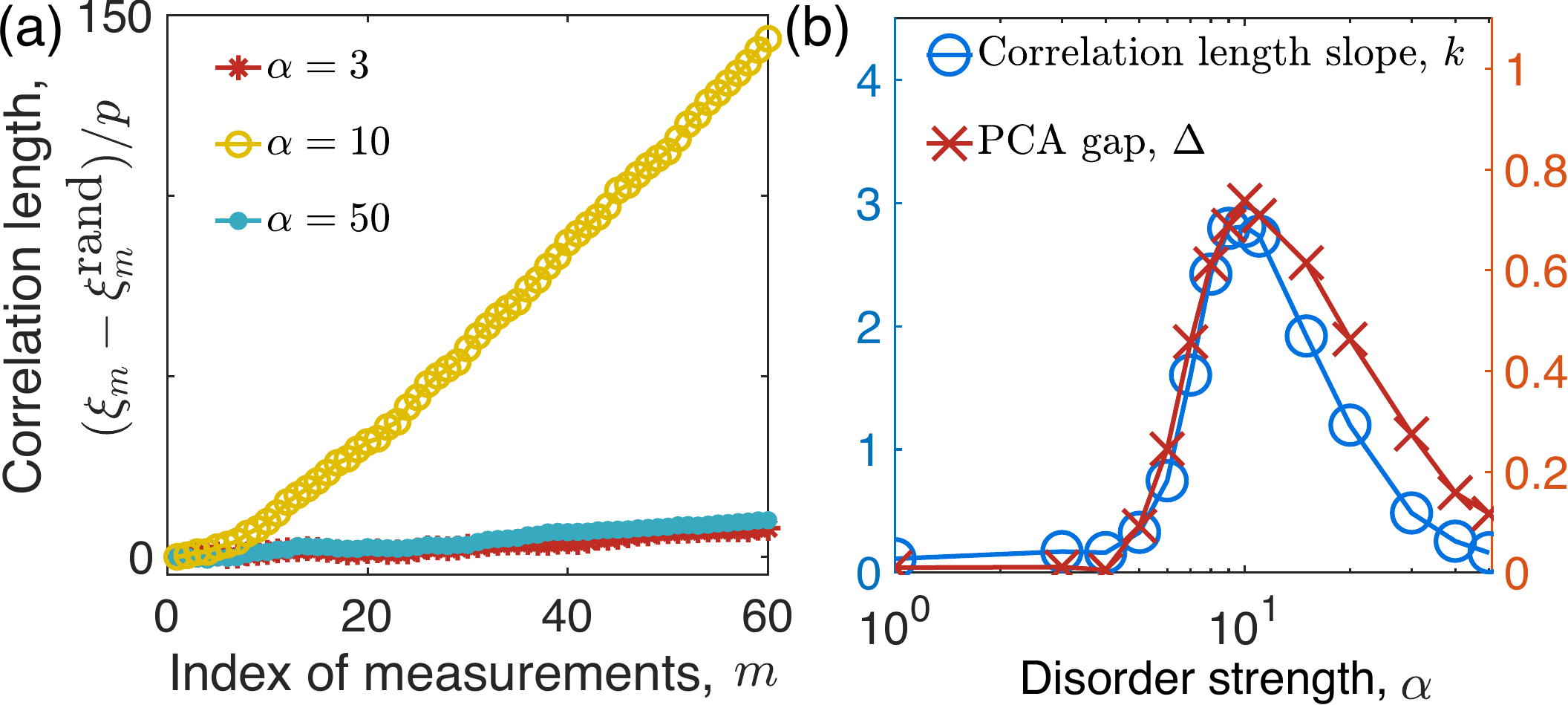}\\
	\caption{(a) Growth of the temporal correlation length $\xi_{m}$ between measurements (normalized by the rate of measurements $p=10^{-4}$) as a function of the number of measurement $m$ for $\alpha=3$, $10$, and $50$. The dashed line shows the $\xi_{m}/p$ for the random dataset. (b) The slope $k$ of the correlation length growth between measurements as a function of disorder strength $\alpha$ compared to the PCA gap $\Delta$.}\label{fig4}
\end{figure}

As a next step, we quantitatively analyze the correlations by computing the ``correlation length" defined as 
\begin{eqnarray}
\xi_{m} = \frac{1}{m} \sum_{i\neq j}^{m} I(i;j),
\label{eq_cl}
\end{eqnarray}
which is actually equal to the average value of the mutual information in the heat map (Fig.~\ref{fig3}(b)) for a smaller square of size $m$. In Fig.~\ref{fig4}(a), 
we plot $(\xi_{m} - \xi^{\text{Rand}}_{m})/p$, where $p = 10^{-4}$ is the rate of measurements, and $\xi^{\text{Rand}}_{m})$ denotes the ``correlation length" calculated by the random dataset, i.e., the elements of (\ref{eq_dataset}) randomly drawn from $\uparrow$ and $\downarrow$, as a function of the index of measurements $m$.

For both the thermal ($\alpha=3$) and deep MBL regime ($\alpha=50$), the behavior of  $\xi_{m}$ is close to the case of random dataset. In contrast, a compellingly more pronounced increase of $\xi_{m}$ is observed in the prethermal MBL regime ($\alpha=10$) indicating a divergent temporal correlation length. As the growth is approximately linear, we extract a correlation length slope $k$ by means of a linear fit, which
is plotted as a function of disorder strength $\alpha$ in Fig.~\ref{fig4}(b).
We find that the nonzero slope $k$ and therefore a divergent correlation length aligns directly with the expected range of the prethermal MBL regime.

Lastly, in Fig.~\ref{fig4}(b), we also plot the PCA gap $\Delta$, i.e., the difference between the largest and second largest PCA eigenvalues. It is shown that the dataset structure identified with the PCA aligns with the observed correlation length slope $k$. 
The resemblance between $\Delta$ and $k$ indicates that the structure of the dataset (\ref{eq_dataset}) in the prethermal MBL regime can be reflected by the strong correlations between the measurement outcomes at different times. 
We also perform the same analysis for the dataset corresponding to larger system sizes, and observe similar phenomena (see Supplementary Materials).

\emph{Discussion and outlook.} -- We show that rare measurements can induce an instability of the prethermal nonergodicity, reflected by the enhancement of EE during monitored dynamics, indicating that rare measurements can be an efficient probe of prethermal nonergodicity. This paves the way to probe other nonergodic quantum phases of matter, such as the Hilbert space fragmentation~\cite{Moudgalya_2022,PhysRevX.10.011047,PhysRevB.101.174204,PhysRevB.110.L220303} and the prethermal dynamics induced by high-frequency periodic driving~\cite{PhysRevResearch.1.033202,PhysRevX.10.011043,PhysRevB.97.245122,PhysRevX.10.021044}, by subjecting rare measurements to the unitary dynamics.  

Furthermore, the measurement outcomes recorded during the dynamics can be employed to learn the properties of the system, which is worthwhile to extend to other non-equilibrium phases of matter, such as conventional MIPTs. Extending the data analysis to include further information from the measurement outcomes, such as the spatial location, into the classical dataset is an intriguing avenue for future research.

~\

The data to generate all figures in this work are available in Zenodo~\cite{heyl_2025_15020002}.

\begin{acknowledgments}
We thank Daniel Braak for insightful discussions.
This project has received funding from the European Research Council (ERC) under the European Union’s Horizon 2020 research and innovation programm (grant agreement No. 853443). 
This work was supported by the German Research Foundation DFG via project 499180199 (FOR 5522).

\end{acknowledgments}

\section*{End matter.}

Here, we provide more details of the calculation of the mutual information $I(i;j)$, which quantifies how much knowledge of the distribution of measurements at time $i$ informs us about the distribution of measurements at time $j$. We first adopt $p_i$ and $p_j$ to represent the probability distributions of measurements at times $i$ and $j$, respectively. Then, we can define the mutual information 
\begin{eqnarray}
I(i;j) = D_{\text{KL}}(p_{ij} || p_{i}p_{j}) = \sum_{p_i,p_j}p_{ij}\log\frac{p_{ij}}{p_{i}p_{j}},
\label{eq_mi}
\end{eqnarray}
which is the Kullback-Leibler divergence between the joint probability distribution of the $i$-th and $j$-th measurement $p_{ij}$, and the product of their marginal distributions $p_{i}$ and $p_{j}$. 

In practice, we estimate $p_{i}$ ($p_{j}$) and $p_{ij}$ by analyzing the columns of the dataset (\ref{eq_dataset}) $\textbf{m}^{(i)}$ $i=1,2,...,N_{\text{sample}}$. For the marginal distributions, we count how many times each outcome appears and normalize the counts to obtain probabilities $p_{X/Y} = \{p_{\downarrow},p_{\uparrow}\}$. Similarly, for the joint probabilities, we count how often each pair of outcomes appears across the samples and normalize these counts to estimate $P_{XY} = \{p_{\uparrow\uparrow},p_{\uparrow\downarrow},p_{\downarrow\uparrow},p_{\downarrow\downarrow}\}$. 

This method, based on simple counting and normalization, provides an empirical estimate of the
probabilities. Importantly, as the number of samples increases, these estimates converge to the
true probabilities according to the law of large numbers.

\bibliography{reference_mipt}

\end{document}


\begin{center}
\textbf{Supplementary Material for ``Probing prethermal nonergodicity through measurement outcomes of monitored quantum dynamics"}
\end{center}

\begin{center}
Zheng-Hang Sun, Fabian Ballar Trigueros, Qicheng Tang, and Markus Heyl
\end{center}

\section{Importance of Born rule}

For the projective measurements in the Z direction $P_{\pm} = (1\pm Z)/2$, we obtain the results in the main text by considering the outcomes of measurements satisfying the Born rule, i.e., the probability of measurements $P_{+}$ ($P_{-}$) for the $i$-th site at a non-equilibrium $|\psi_{n}\rangle$ at time $n$ is $p_{+} = |\langle \psi_{n}  | \uparrow \rangle_{i} |^{2}$ ($p_{-} = |\langle \psi_{n}  | \downarrow \rangle_{i} |^{2}$). 

\begin{figure}[h!]
  \centering
  \includegraphics[width=0.5\linewidth]{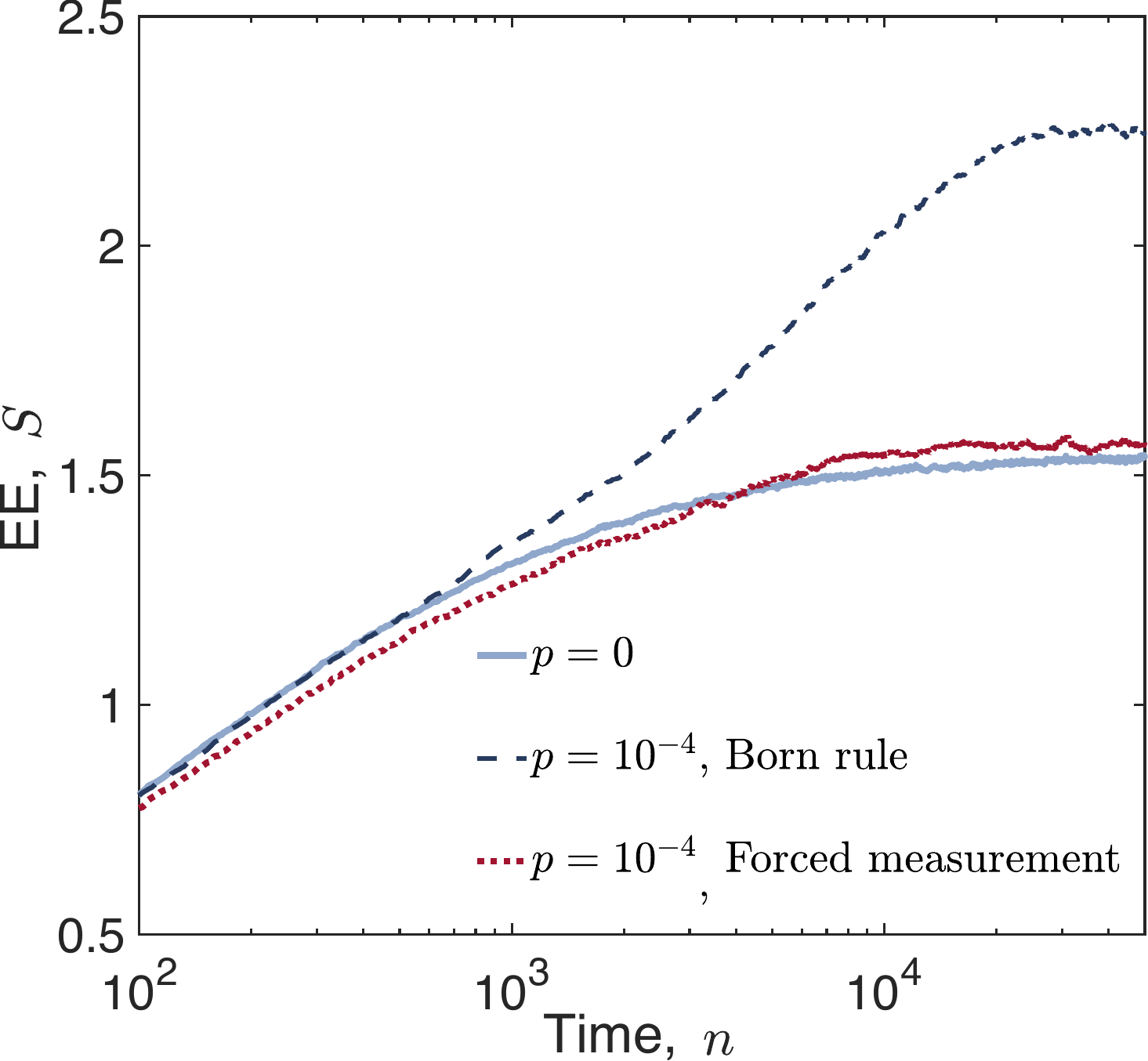}\\
  \caption{  For the quantum circuit in Fig. 1(a) of the main text with the system size $L=12$ and disorder strength $\alpha = 7$, the dynamics of EE for the unitary evolution $p=0$, rare measurements with Born rule and forced rare measurements. For the non-unitary case, the rate of measurements is chosen as $p=10^{-4}$. }\label{s1}
\end{figure}

Here, to show the importance of the Born rule, we consider another case, where all the measurements happen at even (odd) sites are forced to be $P_{-}$ ($P_{+}$), i.e., the  forced measurements. It has been shown that the by subjecting forced measurements to the scrambling unitary dynamics, a disentangled phase can also be induced~\cite{PRXQuantum.2.010352}. In Fig.~\ref{s1}, we plot the dynamics of entanglement entropy (EE) for the hybrid projective-unitary circuits in the Fig.~1(a) of the main text. The system size is $L=12$, and the disorder strength is $\alpha = 7$. We focus on the rare measurements (the rate of measurements $p=10^{-4}$) with both Born rule and forced rule, in comparison with the unitary case $p=0$. One can see that the significant increase of EE is absent for the forced measurements. 

If we consider the forced measurements, one can simply predict that for all the values of $\alpha$, the dataset of  classical measurement outcomes $\textbf{M}$ defined by Eq.~(2) of the main text should be a random dataset, providing no information of the prethermal many-body localized (MBL) regime, which is consistent with the absence of increasing EE in the case of forced measurements (see Fig.~\ref{s1}). This motivates us to gather the information of the prethermal MBL regime from the measurement outcomes based on the Born rule, which plays an important role in the phenomena. In the following parts of the Supplementary Material, we only consider the measurements obeying the Born rule.

\section{Finite-time effect}

\begin{figure}[]
  \centering
  \includegraphics[width=0.8\linewidth]{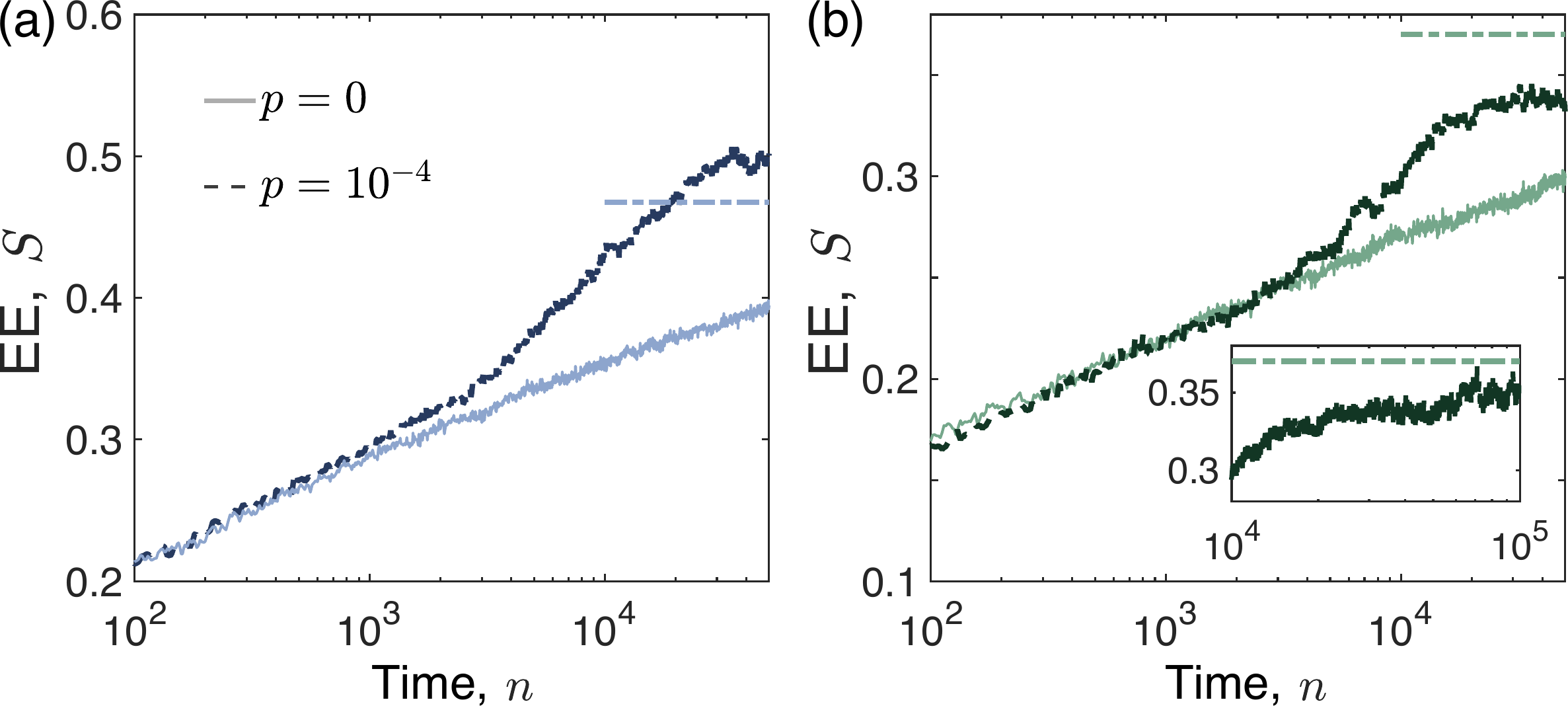}\\
  \caption{  (a) For the quantum circuit in Fig. 1(a) of the main text with the system size $L=12$ and disorder strength $\alpha = 13$, the dynamics of EE for the unitary evolution $p=0$ and rare measurements with a rate $p=10^{-4}$ up to a final time $n=5\times 10^{4}$. (b) is similar to (a) but with the disorder strength $\alpha = 15$. The inset of (b) shows the dynamics up to a longer time $n=10^{5}$. The horizontal lines show the long-time limit of EE with $n=10^{12}$ and $p=0$. }\label{s_a1}
\end{figure}

In this section, we discuss the finite-time effect of the EE dynamics for the monitored quantum circuit with the Floquet model of many-body localization, shown in the Fig.~1(a) of the main text.  In Fig.~\ref{s_a1}, we plot the dynamics of EE for both the unitary ($p=0$) and rare-measurement ($p=10^{-4}$) cases with $\alpha = 13$ and $15$. Here, to study the long-time limit, we compute the EE with $n=10^{12}$ by using exact diagnolization. Even with a finite time $n\sim 5\times 10^{4}$, the EE of rare-measurement case $S(p=10^{-4})$ still exceeds the long-time limit ($n\sim 10^{12}$) of EE for the unitary case [Fig.~\ref{s_a1}(a)]. In contrast, for $\alpha = 15$, although the enhancement of EE with rare measurements is observed at finite time $n\sim 5\times 10^{4}$, we cannot rule out the possibility of $S(p=10^{-4}) \lesssim S(p=0)$ in the limit of $n\rightarrow \infty$.

\begin{figure}[]
  \centering
  \includegraphics[width=0.5\linewidth]{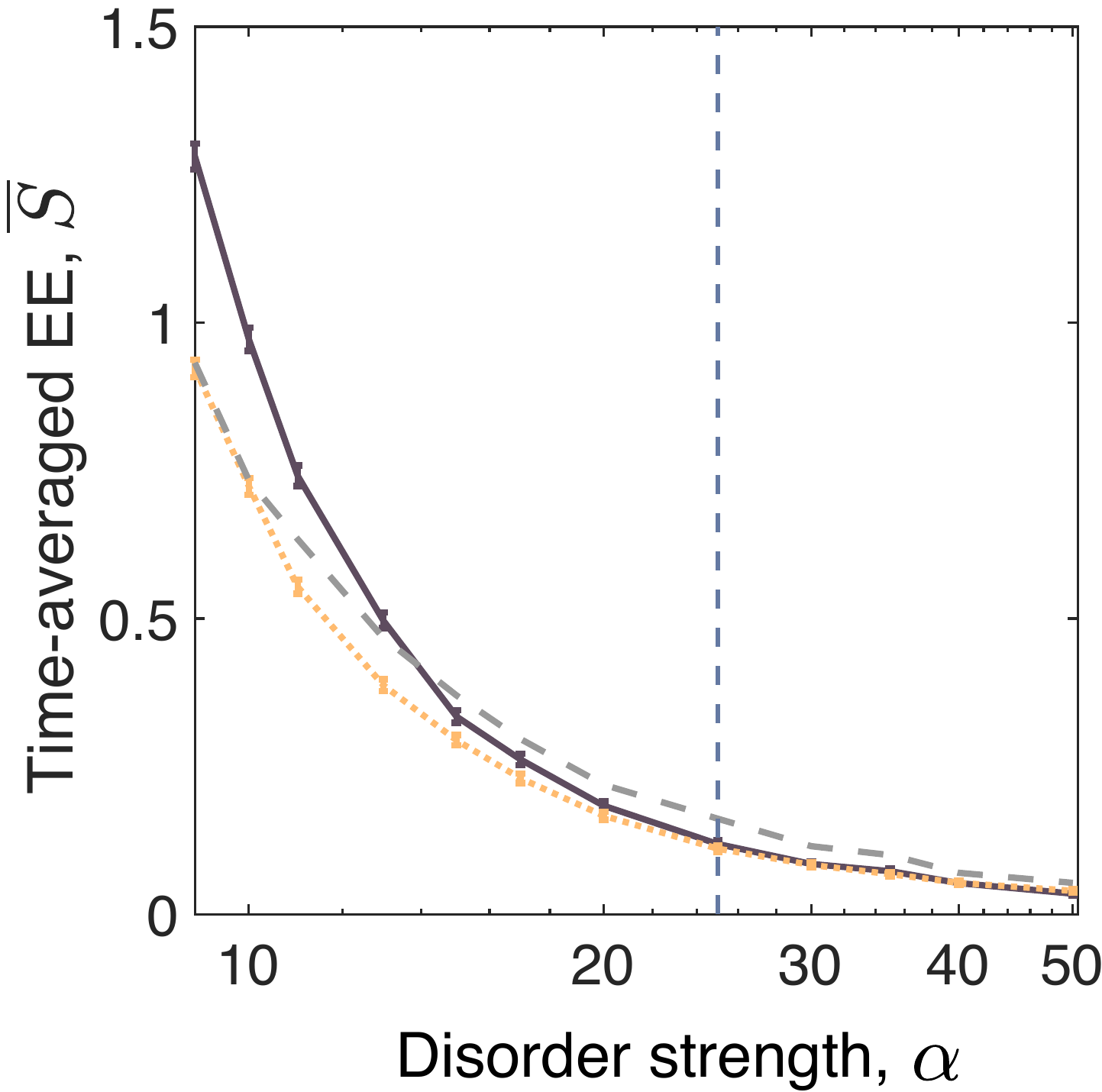}\\
  \caption{ Time-averaged EE as a function of disorder strength $\alpha$. The solid and dotted line shows the time-averaged EE with the time interval $n\in[3\times 10^{4}, 5\times 10^{4}]$ for the rate of measurement $p=10^{-4}$ and the unitary case $p=0$, respectively.  The dashed line shows the long-time limit of EE with $n=10^{12}$ and $p=0$. The vertical dashed line highlights $\alpha=25$.  }\label{s_a2}
\end{figure}

 \begin{figure*}[]
  \centering
  \includegraphics[width=1\linewidth]{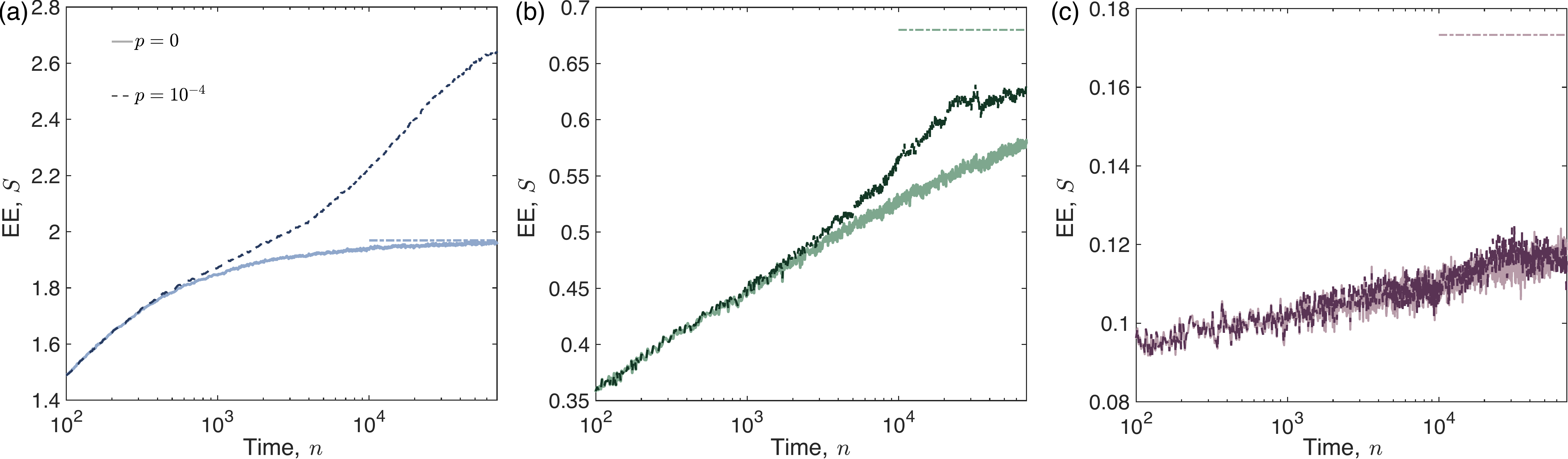}\\
  \caption{ (a) For the dynamics of disordered Heisenberg model (\ref{heisenberg}) with projective measurements, the dynamics of EE of the unitary evolution $p=0$ and rare measurements with a rate $p=2.5\times 10^{-5}$ up to a final time $n=7\times 10^{4}$. Here, the system size is $L=14$, and the disorder strength is $W=3$. (b) is similar to (a) but for $W=7$. (c) is similar to (a) but for $W=20$. The dotted/dashed horizontal lines represent the long-time limit of EE with $n=10^{12}$ and $p=0$.  }\label{s2}
\end{figure*}

In Fig.~\ref{s_a2}, we directly compare the time-averaged EE with an interval $n\in [3\times 10^{4}, 5 \times 10^{4}]$ for both $p=0$ and $p=10^{-4}$, as well as the long-time limit of EE for the unitary case with different disorder strengths $\alpha$. Since with $3\lesssim \alpha \lesssim 13$, $S(p=10^{-4}) > S(p=0)$ even in the infinite-time limit, we argue that $\alpha\simeq 13$ is a lower bound of the finite-size ($L=12$) crossover point between the prethermal MBL and deep MBL regimes. The results in Fig.~\ref{s_a2} also indicates that when $\alpha \gtrsim 25$, $S(p=10^{-4}) \simeq S(p=0)$, and the upper bound of the crossover point can be $\alpha \simeq 25$. Overall, with the system size $L=12$, the crossover point of the deep MBL regime lies in $\alpha \in [13,25]$, which is still larger than the boundary estimated by the level spacing ratio  $\alpha \simeq 6$. 
 
\section{Disordered Heisenberg model}

\begin{figure}[]
  \centering
  \includegraphics[width=0.5\linewidth]{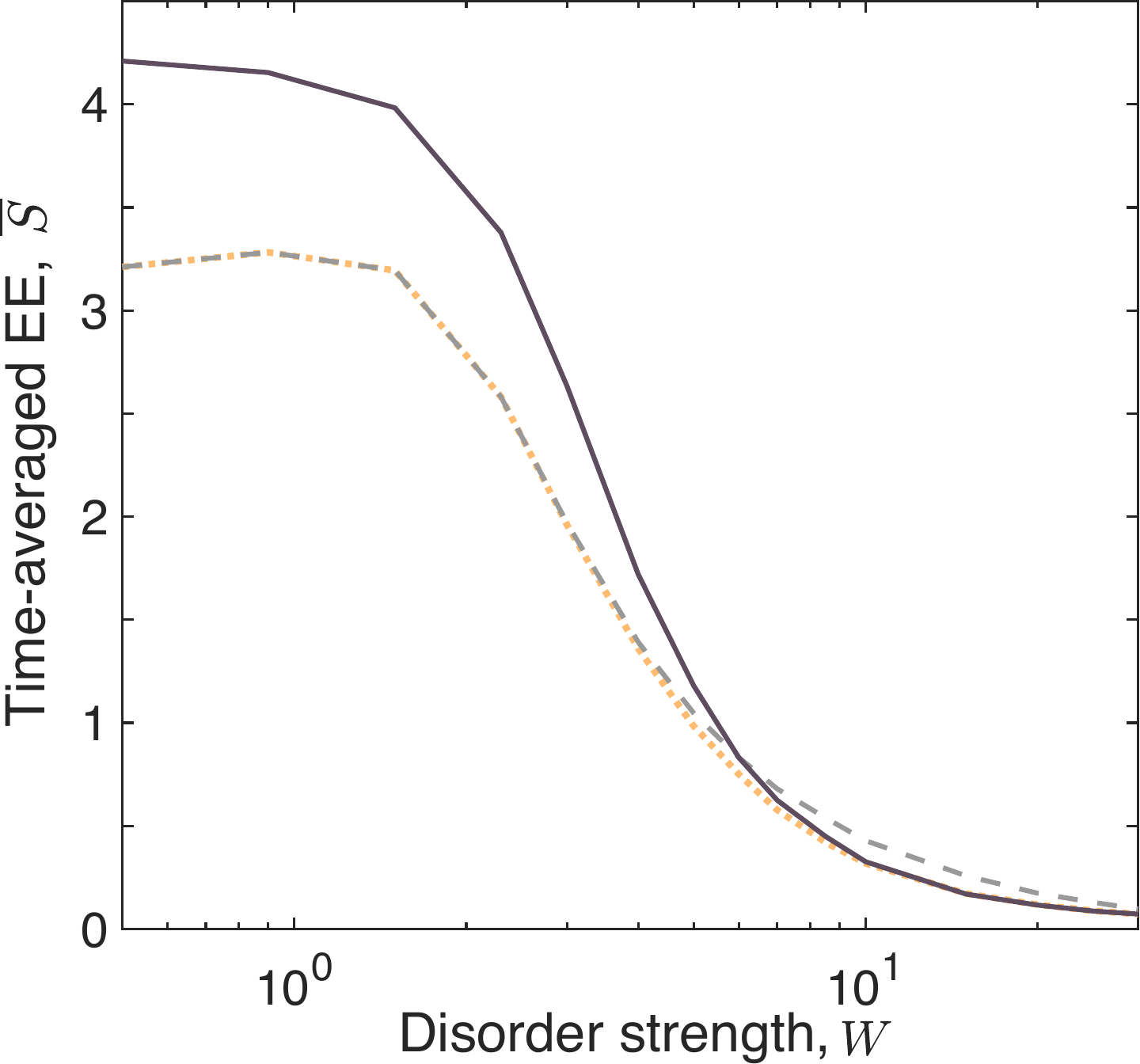}\\
  \caption{Time-averaged EE as a function of disorder strength $W$.  The solid and dotted line shows the time-averaged EE with the time interval $n\in [5\times 10^{4}, 7\times 10^{4}]$ for the rate of measurement $p=2.5\times 10^{-5}$ and the unitary case $p=0$, respectively. The dashed line shows the long-time limit of EE with $p=0$ and $n=10^{12}$. For the system size $L=14$, the Page value (\ref{page}) is $S_{P}\simeq 4.35$, which is much larger than the EE of unitary dynamics with $W\simeq 0$. }\label{s3}
\end{figure}

In the main text, we mainly focus on the random-circuit model of many-body localization. In this section, we study whether the rare measurement can capture the prethermal MBL regime on the disorder Heisenberg model, as a typical Hamiltonian model of many-body localization. 

The Hamiltonian of the disordered Heisenberg model can be written as 
\begin{equation}
    \hat{H} = \sum_{i=1}^{L-1}(\hat{S}_{i}^{x}\hat{S}_{i+1}^{x} +\hat{S}_{i}^{y}\hat{S}_{i+1}^{y} +\hat{S}_{i}^{z}\hat{S}_{i+1}^{z}  ) + \sum_{i=1}^{L} h_{i} \hat{S}_{i}^{z},
    \label{heisenberg}
\end{equation}
where $S_{i}^{\alpha}$ ($\alpha\in\{x,y,z\}$) represents the spin-1/2 operators in the $i$-th site, $L$ denotes the system size, and $h_{i}$ is randomly drawn from the uniform distribution $[-W,W]$, with $W$ being the disorder strength. Previous studies of the model (\ref{heisenberg}) based on the conventional landmarks of many-body localization transitions, such as the level spacing ratio, indicate that the crossover point is $W\simeq 3$~\cite{PhysRevB.105.174205}. 

Here, we study the dynamics of entanglement entropy for the hybrid projective-unitary circuits with the same structure in Fig.~1(a) of the main text but replacing the unitary part by $\hat{U} = \exp(-i\hat{H})$, where the Hamiltonian $\hat{H}$ is given by Eq.~(\ref{heisenberg}). The initial state is chosen as the N\'{e}el state. The system size is $L=14$, and the rate of rare measurements is $p=2.5\times 10^{-5}$. 

In Fig.~\ref{s2}, we plot the dynamics of EE for three values of disorder strength $W=3$, $10$, and $20$. With $W=3$, we observe that there is an obvious increase of EE induced by rare measurements, which is even larger than the unitary case in the infinite-time limit. With $W=10$, there is an enhancement of entanglement in the finite time. With $W=20$, the increase of EE is absent, and the system lies in a stable MBL regime. 

More systematic results are shown in Fig.~\ref{s3}, where we directly compare the time-averaged EE with an interval $n\in [5\times 10^{4}, 7\times 10^{4}]$ for both $p=0$ and $p=2.5\times 10^{-5}$, as well as the long-time limit of EE for the unitary case with different disorder strengths $W$. With $W \lesssim 6$, $S(p=2.5\times 10^{-5}) > S(p=0)$ is satisfied even in the infinite-time limit, and when $W \gtrsim 15$, $S(p=2.5\times 10^{-5}) \simeq S(p=0)$ for finite time dynamics. Thus, we estimate the  lower and upper bound of the crossover point between the prethermal MBL and deep MBL regime as $W\simeq 6$ and $W\simeq 15$, respectively. 

We note that, different from the random-circuit Floquet model of many-body localization studied in the main text, for the disordered Heisenberg model with small $W$, there is no well-thermalizing phase which is robust against rare measurements. Actually, the value of EE when $W\simeq 0$ is smaller than the Page value of half-chain EE defined by 
\begin{equation}
    S_{P}=0.5[L\log(2) - 1]
    \label{page}
\end{equation}
with $L$ being the system size. 

\section{Influence of the number of samples}

\begin{figure}[]
  \centering
  \includegraphics[width=0.5\linewidth]{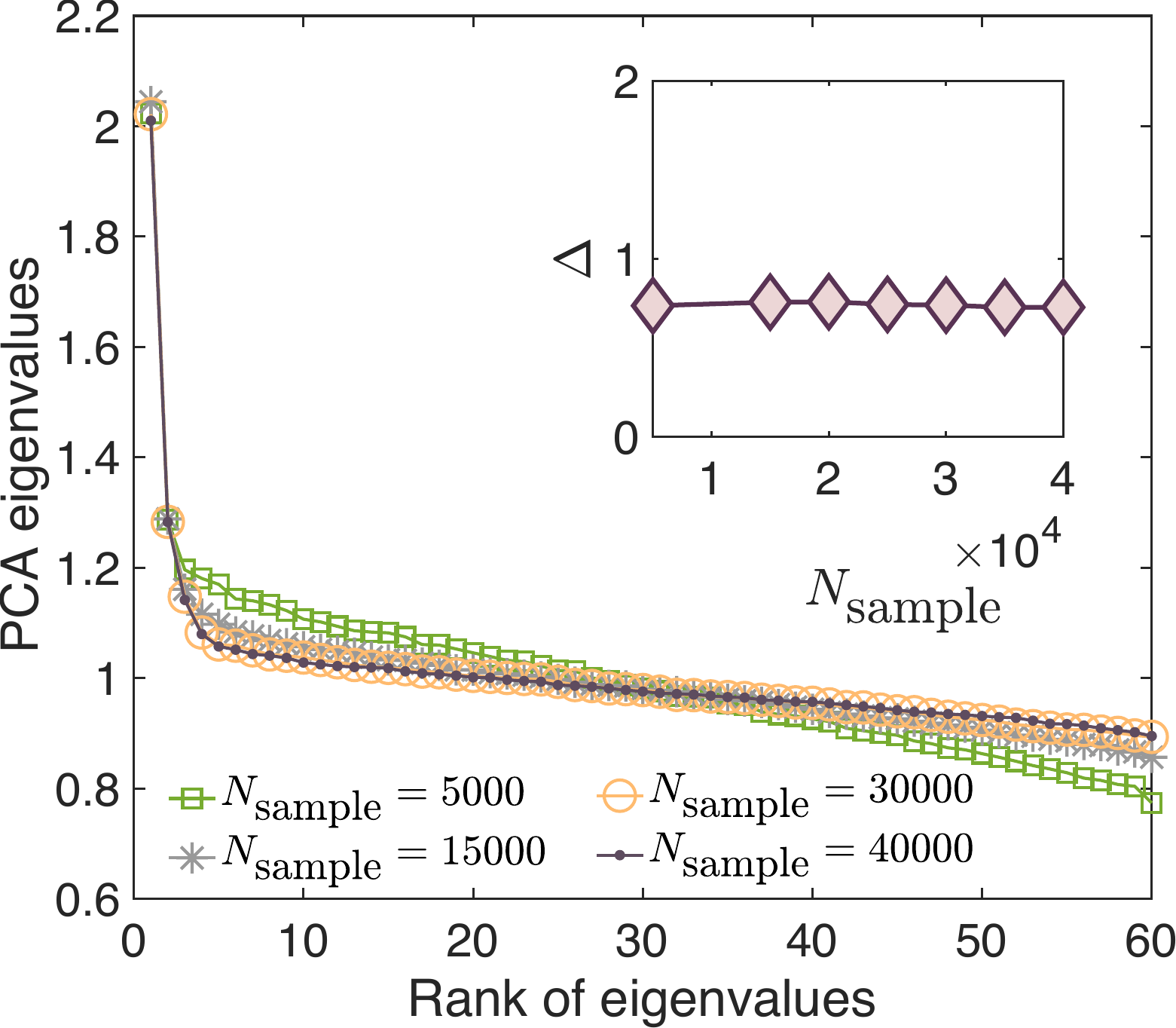}\\
  \caption{The PCA eigenvalues for the dataset extracted from the dynamics of the \textcolor{red}{hybird} quantum circuit, with the system size $L=12$ and disorder strength $\alpha = 10$. Here, we consider several numbers of samples $N_{\text{sample}} =5\times 10^{3}$, $1.5\times 10^{4}$, $3\times 10^{4}$, and $4\times 10^{4}$. The inset shows the difference between the largest and second largest PCA eigenvalue $\Delta$ as a function of $N_{\text{sample}}$. }\label{s4}
\end{figure}

In this section, we discuss the influence of the number of samples $N_{\text{sample}}$ on the analysis of classical dataset consisting of measurement outcomes, i.e., the dataset $\textbf{M}$ defined in Eq.~(2) in the main text with different  $N_{\text{sample}}$. 

We first study the PCA eigenvalues with different $N_{\text{sample}}$. Here, we consider the system size $L=12$, and $\alpha=10$ as a typical example. The results are shown in Fig.~\ref{s4}. One can see that the behavior of PCA eigenvalues as a function of the rank of eigenvalues saturates with $N_{\text{sample}} \gtrsim 3\times 10^{4}$. We also plot the difference between the largest and second largest PCA eigenvalue $\Delta$, as a key quantification of the effective dimension of the dataset $\textbf{M}$, as a function of $N_{\text{sample}}$, in the inset of Fig.~\ref{s4}. Although the small PCA eigenvalues are sensitive to the $N_{\text{sample}}$, the large PCA eigenvalues and $\Delta$ are robust to the $N_{\text{sample}}$. 

\begin{figure}[]
  \centering
  \includegraphics[width=1\linewidth]{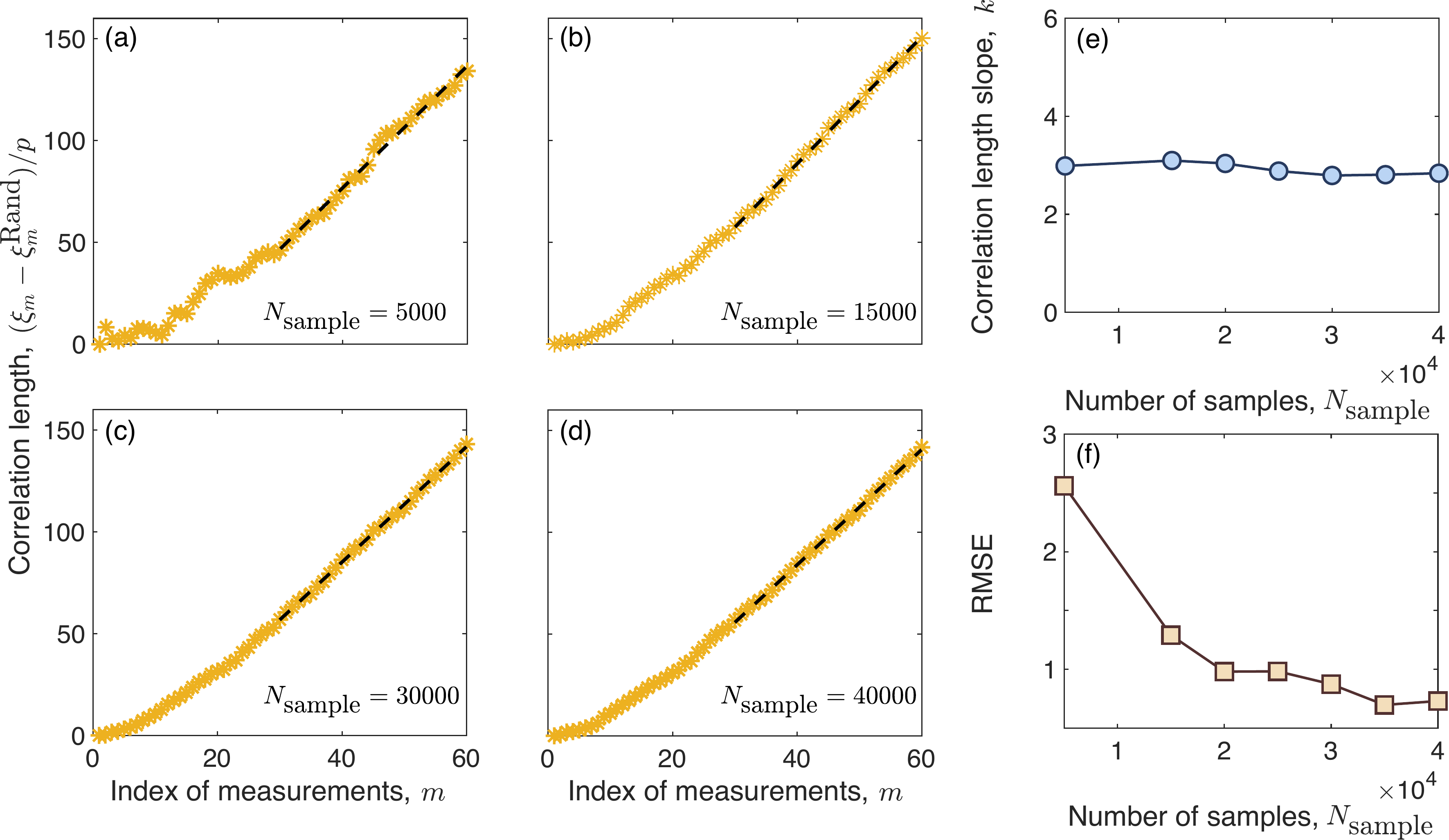}\\
  \caption{(a) The ``correlation length" $(\xi_{m} - \xi_{m}^{\text{Rand}})/p$ for different index of measurements $m$ with system size $L=12$, disorder strength $\alpha=10$, and the number of samples $N_{\text{sample}}=5000$. (b) is similar to (a) but for $N_{\text{sample}}=15000$. (c) is similar to (a) but for $N_{\text{sample}}=30000$. (d) is similar to (a) but for $N_{\text{sample}}=40000$. The dashed lines in (a)-(d) are the linear fitting of the data $(\xi_{m} - \xi_{m}^{\text{Rand}})/p$ with $m\in[30,60]$. (e) The slope obtained by the linear fitting $k$ as a function of $N_{\text{sample}}$. (f)  The root mean squared error (RMSE) for the linear fitting as a function of $N_{\text{sample}}$. }\label{s5}
\end{figure}

We then focus on the ``correlation length" $\xi_{m}$ defined by Eq.~(4) of the main text. In Fig.~\ref{s5}(a)-(d), we plot the $(\xi_{m} - \xi_{m}^{\text{Rand}})/p$, with $\xi_{m}^{\text{Rand}}$ being the ``correlation length" for the random dataset, and $p=10^{-4}$ being the rate of measurements, as a function of the index of measurements $m$. It is seen that with a relatively small number of samples, i.e., $N_{\text{sample}} = 5000$, $(\xi_{m} - \xi_{m}^{\text{Rand}})/p$ exhibits an overall linear growth trend, albeit with fluctuations around the ideal linear behavior. With the increase of $N_{\text{sample}}$, it becomes more perfectly linear [also see Fig.~\ref{s5}(f) for the root mean squared error (RMSE) for the linear fitting]. Moreover, we plot the correlation length slope $k$ obtained by fitting the data of $(\xi_{m} - \xi_{m}^{\text{Rand}})/p$ with different $N_{\text{sample}}$ in Fig.~\ref{s5}(e). We show that, similar to the $\Delta$, the slope $k$ is robust to the $N_{\text{sample}}$.

\section{Results for systems with larger sizes}

All the results in the main text are for the hybrid quantum circuits with the system size $L=12$. In this section, we show that the phenomena observed for the systems with $L=12$ still exist for larger system size. For larger system sizes $L=14$ and $16$, we first display the data of EE for larger system sizes, and then we present the results of the analysis of classical dataset. Finally, we give some additional discussions and outlooks for the finite-size effects.

\subsection{Entanglement entropy}

\begin{figure}[]
  \centering
  \includegraphics[width=0.66\linewidth]{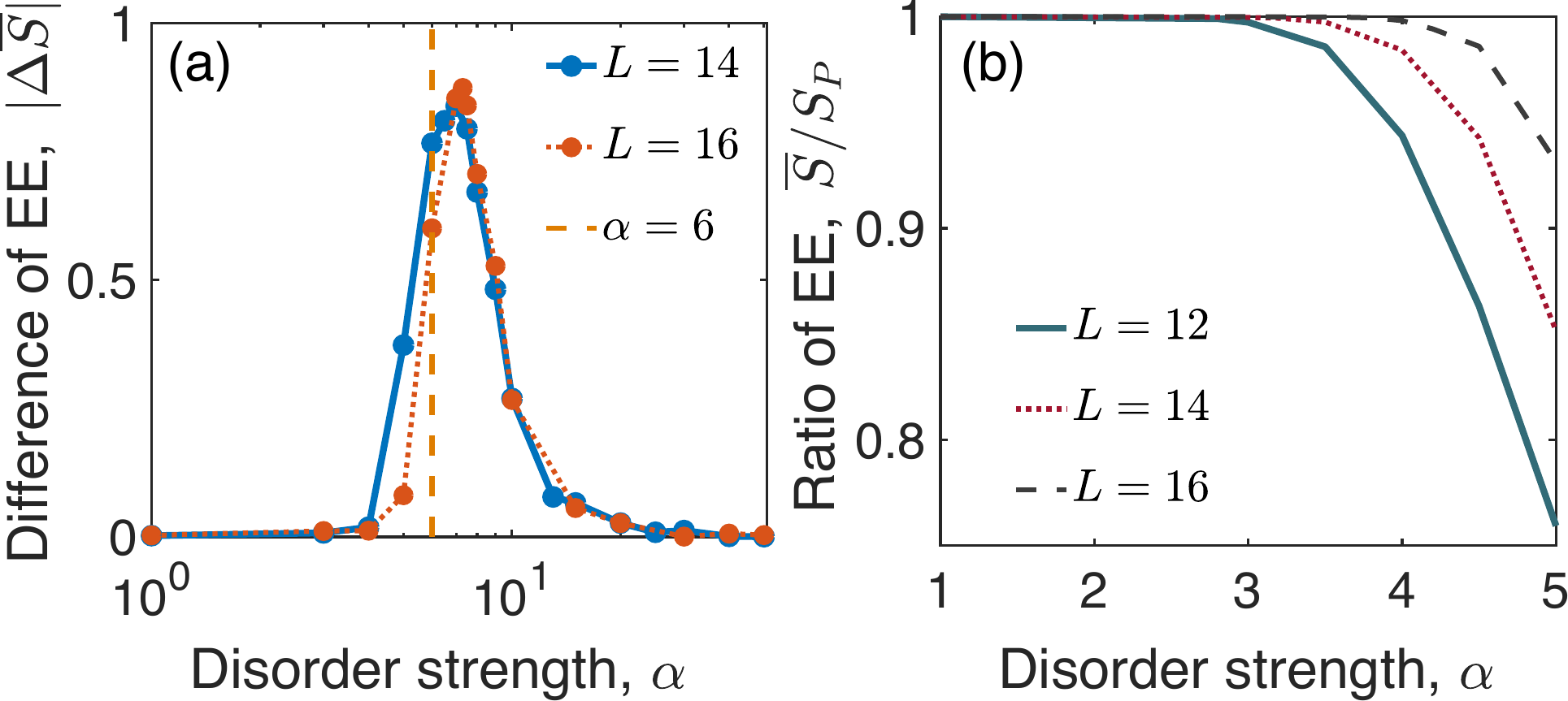}\\
  \caption{(a) For system sizes $L=14$ and $16$, the difference of long-time averaged EE between the unitary case ($p=0$) and the quantum circuit with rare measurements ($p=10^{-4}$), i.e., $|\Delta \overline{S}| = |\overline{S}(p=10^{-4}) - \overline{S}(p=0)|$ as a function of disorder strength $\alpha$. The vertical dashed line highlights $\alpha=6$, as the maximum point of  $|\Delta \overline{S}|$ for the data with the system size $L=12$ [see Fig.~1(c) of the main text]. (b) For the unitary quantum circuit with $p=0$ and different system sizes, the ratio of EE $\overline{S}/S_{P}$, where $\overline{S}$ denotes the long-time averaged EE, and $S_{P}$ is the Page value, as a function of $\alpha$.}\label{s6}
\end{figure}

\begin{figure}[]
  \centering
  \includegraphics[width=0.7\linewidth]{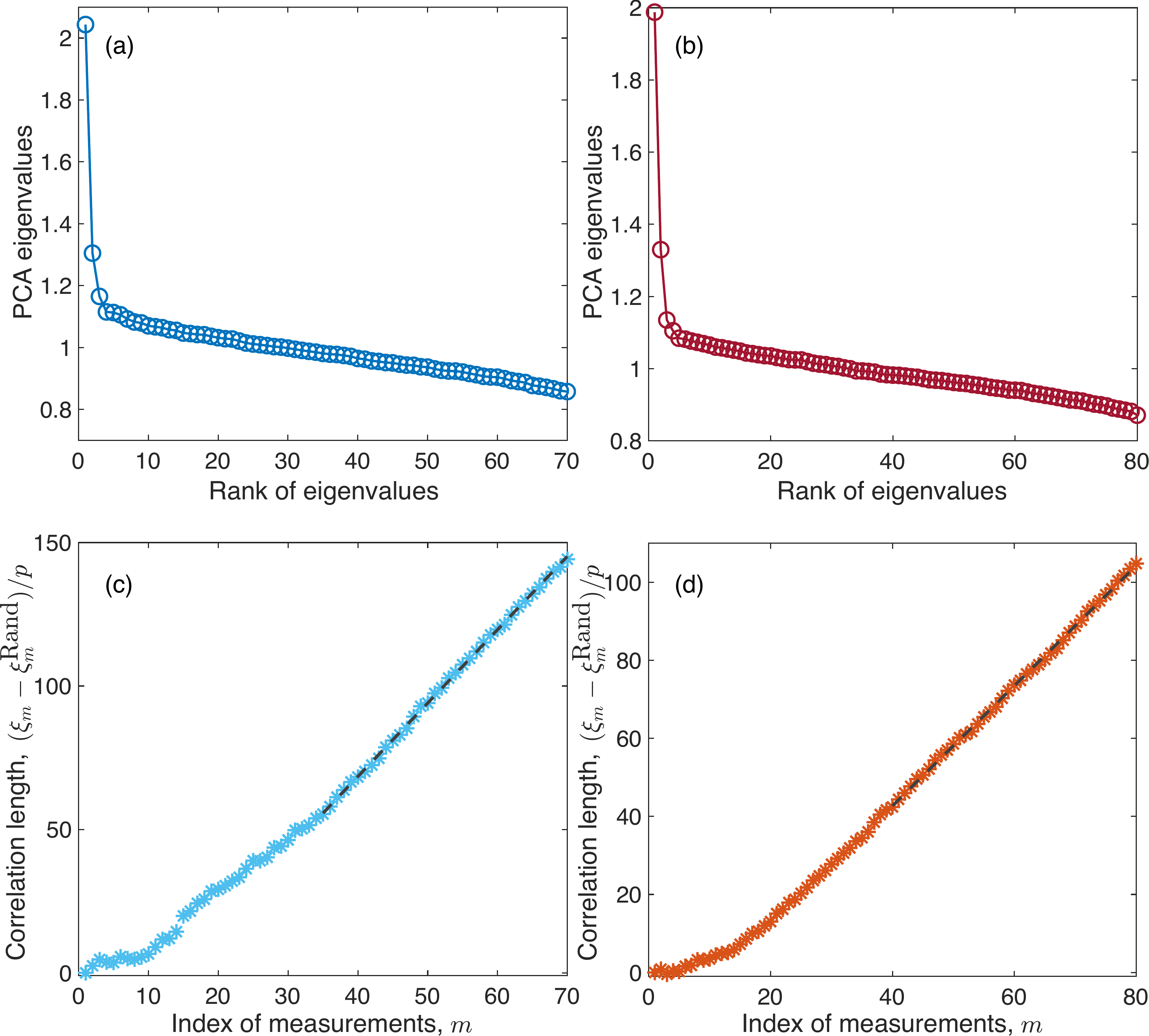}\\
  \caption{(a) The PCA eigenvalues for the dataset extracted from the dynamics of the quantum circuit with system size $L=14$, disorder strength $\alpha = 10$, and the number of samples $N_{\text{sample}}=30000$. (b) is similar to (a) but for the system size $L=16$. (c) The ``correlation length" $(\xi_{m} - \xi_{m}^{\text{Rand}})/p$ for different index of measurements $m$ with system size $L=14$,  disorder strength $\alpha = 10$, and the number of samples $N_{\text{sample}}=30000$. (d) is similar to (c) but for the system size $L=16$. The dashed lines in (c) and (d) are the linear fitting of the data $(\xi_{m} - \xi_{m}^{\text{Rand}})/p$ with $m\in[35,70]$ and $m\in[40,80]$, respectively.}\label{s7}
\end{figure}

In Fig.~\ref{s6}(a), we show the difference of time-averaged EE $|\Delta \overline{S}| = |\overline{S}(p=10^{-4}) - \overline{S}(p=0)|$ for the hybrid projective-unitary circuit in the Fig.~1(a) of the main text with larger system sizes $L=14$ and $16$. We also highlight the location of the maximum point of   $|\Delta \overline{S}|$ for the system size $L=12$, i.e., $\alpha = 6$ (see the Fig.~1(c) of the main text). One can see that with increasing system sizes $L$, the the location of the maximum point of   $|\Delta \overline{S}|$ tends to a larger value. This can be qualitatively explained by the asymptotic features of the thermal phase. In Fig.~\ref{s6}(b), we plot the ratio between the time-averaged EE and the Page value, i.e.,  $\overline{S} / S_{P}$, as a function of $\alpha$, showing that the boundary between the thermal phase with $\overline{S} / S_{P} \simeq 1$, and the prethermal MBL regime with $\overline{S} / S_{P} < 1$, also tends to larger values.

\subsection{Analysis of classical dataset}

Based on the results shown in Fig.~\ref{s5}, the number of samples $N_{\text{sample}} = 30000$ is sufficient to obtain saturated results of PCA and mutual information. Here, we consider larger system sizes $L=14$ and $16$. Similar to the case of the system size $L=12$, with the rate of measurements $p=10^{-4}$, we can fix the number of measurements $M=70$ and $80$ for $L=14$ and $16$, respectively. 

The results of PCA for larger system sizes $L=14$ and $16$ are shown in Fig.~\ref{s7}(a) and (b), respectively. We also display the quantity $(\xi_{m} - \xi_{m}^{\text{Rand}})/p$ [the same quantity shown in Fig.~\ref{s5}(a)-(d)]. In short, with the disorder strength $\alpha=10$, as a typical value in the prethermal MBL regime, the behaviors of PCA eigenvalues and the ``correlation length" are similar to the case of $L=12$ shown in the main text.


\subsection{Additional discussions of finite-size effect}

Whether the MBL regime can be a stable dynamical phase of matter in the thermodynamic limit is a challenging problem 
due to the presence of strong finite-size effects~\cite{Sierant_2025}. Recent studies of characterizing the onset of MBL regime by considering the Berezinskii–Kosterlitz–Thouless scaling suggest that in the thermodynamic limit, the models of many-body localization remains ergodic at any disorder strength~\cite{PhysRevB.102.064207,PhysRevE.102.062144}. 

Although finite-size analysis is not the primary focus of this work, in Fig.~\ref{s6}, the extension of the thermal phase and the broad prethermal MBL regime quantified by the instability against rare measurements in larger system sizes indicate that for the accessible system sizes, there exists only a deep MBL regime instead of a MBL phase. 

For further explorations, it is worthwhile to study the dynamics of the models of many-body localization with rare measurements for larger system sizes by using matrix-product-state based algorithms. Such approaches could provide deeper insights into the stability of MBL regimes as they approach the thermodynamic limit.

\bibliography{reference_mipt}